\pdfoutput=1

\documentclass[twocolumn,prd,reprint,preprintnumbers,amsmath,amssymb,superscriptaddress,nofootinbib]{revtex4}

\usepackage{aas_macros}
\usepackage{enumerate}
\usepackage{epsfig,psfrag,graphics,verbatim}
\usepackage{graphicx}
\usepackage{dcolumn}
\usepackage{bm}
\usepackage{color}

\usepackage{multirow}

\begin{document}
\preprint{TUM-HEP 993/15}

\title{Dynamical constraints on the dark matter distribution in the Milky Way}

\author{Miguel Pato}
\affiliation{The Oskar Klein Centre for Cosmoparticle Physics, Department of Physics, Stockholm University, AlbaNova, SE-106 91 Stockholm, Sweden}
\affiliation{Physik-Department T30d, Technische Universit\"at M\"unchen, James-Franck-Stra\ss{}e, D-85748 Garching, Germany}

\author{Fabio Iocco}
\affiliation{ICTP South American Institute for Fundamental Research, and Instituto de F\'isica Te\'orica - Universidade Estadual Paulista (UNESP), Rua Dr.~Bento Teobaldo Ferraz 271, 01140-070 S\~{a}o Paulo, SP Brazil}
\affiliation{Instituto de F\'isica Te\'orica UAM/CSIC, C/ Nicol\'as Cabrera 13-15, 28049 Cantoblanco, Madrid, Spain}

\author{Gianfranco Bertone}
\affiliation{GRAPPA Institute, University of Amsterdam, Science Park 904, 1090 GL Amsterdam, The Netherlands}

\date{\today}

\begin{abstract}
An accurate knowledge of the dark matter distribution in the Milky Way is of crucial importance for galaxy formation studies and current searches for particle dark matter. In this paper we set new dynamical constraints on the Galactic dark matter profile by comparing the observed rotation curve, updated with a comprehensive compilation of kinematic tracers, with that inferred from a wide range of observation-based  morphologies of the bulge, disc and gas. The generalised Navarro-Frenk-White (NFW) and Einasto dark matter profiles are fitted to the data in order to determine the favoured ranges of local density, slope and scale radius. For a representative baryonic model, a typical local circular velocity $v_0=230\,$km/s and a distance of the Sun to the Galactic centre $R_0=8\,$kpc, we find a local dark matter density $\rho_0 = 0.420^{+0.021}_{-0.018} \, (2\sigma) \pm 0.025 \,\textrm{GeV/cm}^3$ ($\rho_0 = 0.420^{+0.019}_{-0.021} \, (2\sigma) \pm 0.026 \,\textrm{GeV/cm}^3$) for NFW (Einasto), where the second error is an estimate of the systematic due to baryonic modelling. Apart from the Galactic parameters, the main sources of uncertainty inside and outside the solar circle are baryonic modelling and rotation curve measurements, respectively. Upcoming astronomical observations are expected to reduce all these uncertainties substantially over the coming years.
\end{abstract}

\maketitle

\section{Introduction} 

\par The observation of flat rotation curves in spiral galaxies was historically one of the cornerstones upon which the case for dark matter was built (e.g.~\cite{1970ApJ...160..811F,1973A&A....26..483R,1978PhDT.......195B,1978ApJ...225L.107R,1979ARA&A..17..135F}; for more recent studies, see e.g.~\cite{Persic:1995ru,Salucci:2007tm,2009MNRAS.397.1169D,2009Natur.461..627G}). In our Galaxy, a spiral itself, measuring the rotation curve remains however a daunting task due to our position inside the stellar disc. As a consequence, the knowledge of the Galactic distribution of dark matter is affected by sizeable uncertainties, especially in the inner regions, where baryons dominate the gravitational potential. This unfortunate circumstance hinders the study of the formation history of the Milky Way and the interpretation of particle dark matter searches (see e.g.~\cite{Bertone:2010zza,Jungman:1995df,Bergstrom00,Bertone05}).

\par Several techniques have however been designed to provide gravitational constraints on the dark matter distribution in the Galaxy, and in particular in the solar neighbourhood. They fall into two classes: local methods \cite{,,,Smith:2011fs,Garbari:2012ff,Zhang:2012rsb,Read:2014qva,2012ApJ...751...30M,Bovy:2012tw,2015A&A...573A..91M} (see also \cite{Bovy:2013raa,2014ApJ...794..151L}), based on a Jeans-Poisson analysis of the kinematics of local stars, and global methods \cite{Dehnen:1996fa,Sofue2009,CatenaUllio2010,Weber:2009pt,Salucci:2010qr,2010ApJ...720L.108G,Iocco2011,2011MNRAS.414.2446M,Nesti:2013uwa,2015arXiv150405368S}, which involve the mass modelling of the Galaxy as a whole and use global dynamical constraints. Each class of methods brings complementary information on the dark matter distribution and both suffer from systematics. For example, the precise measurements sometimes provided by global methods come at the expense of enforcing a preassigned shape to the dark matter profile in the inner Galaxy, where the effect of baryons is potentially large. This in turn leads to sizeable systematics in the determination of the dark matter profile due to the uncertainties on the distribution of baryons.

\par In Ref.~\cite{Iocco2011}, we have applied a global method to show that interesting constraints on the dark matter distribution in our Galaxy can be achieved by comparing the rotation curve inferred from baryonic models with actual rotation curve measurements. In light of our recent work \cite{2015NatPh..11..245I}, here we set out to revisit the previous analysis in Ref.~\cite{Iocco2011} and improve it with three key ingredients:
\begin{enumerate}[i)]
\item a new and more complete compilation of rotation curve data, including 2780 measurements from gas kinematics, star kinematics and masers (Sec.~\ref{secobs});
\item a wide range of baryonic models, including virtually all data-based morphologies for the bulge, disc and gas (Sec.~\ref{secobs}); and
\item a procedure based on a two-dimensional $\chi^2$ applied to unbinned data (Sec.~\ref{secmethod}).
\end{enumerate}

\par The results, which supersede those of our previous work \cite{Iocco2011}, are presented in Sec.~\ref{secDM} in the form of constraints on the dark matter profile. Our conclusions are given in Sec.~\ref{secconc}.

\section{Observations}\label{secobs}

\par Our analysis of the dark matter distribution in the inner Galaxy relies upon two observational inputs: the rotation curve and the spatial distribution of stars and gas. Let us begin with the former. Rotation curve measurements in the Milky Way have been available for several decades now, but the corresponding literature is rather scattered. In Ref.~\cite{2015NatPh..11..245I} we have assembled a new, comprehensive compilation of kinematic tracers across the Galaxy that comprises:
\begin{itemize}
\item \emph{gas kinematics}, including HI terminal velocities \cite{Fich1989,Malhotra1995,McClure-GriffithsDickey2007}, HI thickness \cite{HonmaSofue1997}, CO terminal velocities \cite{BurtonGordon1978,Clemens1985,Knapp1985,Luna2006}, HII regions \cite{Blitz1979,Fich1989,TurbideMoffat1993,BrandBlitz1993,Hou2009} and giant molecular clouds \cite{Hou2009};
\item \emph{star kinematics}, including open clusters \cite{FrinchaboyMajewski2008}, planetary nebulae \cite{Durand1998}, classical cepheids \cite{Pont1994,Pont1997} and carbon stars \cite{DemersBattinelli2007,Battinelli2013}; and
\item \emph{masers} \cite{Reid2014,Honma2012,StepanishchevBobylev2011,Xu2013,BobylevBajkova2013}.
\end{itemize}
This literature survey is particularly exhaustive in the range of Galoctocentric radii $R=3-20\,$kpc. It intentionally leaves out tracers with relevant random motions, asymmetric drift or kinematic distances only in an effort to track Galactic rotation as reliably as possible. From the references listed above we additionally exclude individual objects close to the Galactic centre or anti centre, with incomplete data or in some way signalled as suspicious. The final compilation consists of 2780 objects across the range $R=0.5-25\,$kpc, of which 2174, 506 and 100 are contributed by gas kinematics, star kinematics and masers, respectively. Each object is characterised by its Galactic coordinates $(\ell,b)$, heliocentric distance $d$, Galactocentric radius $R=(d^2 \cos^2 b + R_0^2 - 2 R_0 d \cos b \, \cos \ell)^{1/2}$ (with $R_0$ the distance of the Sun to the Galactic centre) and line-of-sight velocity in the local standard of rest (LSR) $v_{\textrm{lsr}}^{\textrm{los}}$. Uncertainties on $d$ and $v_{\textrm{lsr}}^{\textrm{los}}$ are carefully extracted or estimated from the original references, whereas those on $\ell$ and $b$ are neglected since they are by far sub-dominant in all instances. The angular velocity $\omega_c(R)$ and corresponding uncertainties are then found by inverting (and propagating) the well-known expression
\begin{equation}\label{vlos}
v_{\textrm{lsr}}^{\textrm{los}} = \left( R_0\omega_c - v_0  \right) \cos b \, \sin \ell \, ,
\end{equation}
where $v_0\equiv v_c(R_0)$ is the local circular velocity. Note that we use $\omega_c$ instead of $v_c\equiv R \omega_c$ to avoid the correlation between the errors of $R$ and $v_c$. In the end, given $R_0$, $v_0$ and a local standard of rest, each object provides an independent constraint on $R$ and $\omega_c(R)$. Throughout the work, unless otherwise specified, we use $R_0=8\,$kpc, $v_0=230\,$km/s and the peculiar solar motion $\left(U,V,W\right)_{\odot}=(11.10,12.24,7.25)\,$km/s \cite{Schoenrich2010}. The release of a user-friendly tool to retrieve all kinematic data used here is planned for the near future \cite{IoccoPatoTool}.

\begin{table}
\begin{center}
\begin{tabular}{ |l| ccc l cc|	} 
\cline{2-7}
 \multicolumn{1}{c|}{} &	 & model & & specification && Ref.\\
\hline
\multirow{7}{*}{bulge} 	& & $\,\,\,$1$\color{black}^{\ast}$& & 	exponential E2			&& \cite{Stanek1997}\\
			& & 2 & & 				gaussian G2 			&& \cite{Stanek1997}\\
			& & 3 & & 				gaussian plus nucleus 		&& \cite{Zhao1996}\\
			& & 4 & & 				truncated power law		&& \cite{BissantzGerhard2002}\\
			& & 5 & & 				power law plus long bar		&& \cite{LopezCorredoira2007}\\
			& & 6 & & 				truncated power law		&& \cite{Vanhollebeke2009}\\
			& & 7 & & 				double ellipsoid 		&& \cite{Robin2012}\\
\hline
\multirow{5}{*}{disc} 	& & 1 & & 				thin plus thick 		&& \cite{HanGould2003}\\
			& & 2 & & 				thin plus thick 		&& \cite{CalchiNovatiMancini2011}\\
			& & 3 & & 				thin plus thick plus halo	&& \cite{deJong2010}\\
			& & 4 & & 				thin plus thick plus halo	&& \cite{Juric2008}\\
			& & $\,\,\,$5$\color{black}^{\ast}$ & & single maximal disc		&& \cite{Bovy:2013raa}\\
\hline
\multirow{2}{*}{gas} 	& & $\,\,\,$1$\color{black}^{\ast}$ & & H$_2$, HI, HII			&& \cite{Ferriere1998}\\
			& & 2 & & 				H$_2$, HI, HII 			&& \cite{Moskalenko2002}\\	
\hline
\end{tabular}
\caption{Summary of all models of stellar bulge, stellar disc and gas used to describe the baryonic component of our Galaxy. For further details, please see Ref.~\cite{2015NatPh..11..245I}. The configurations defining the representative baryonic model used later on in our analysis are indicated with an asterisk.}\label{tab:models}
\end{center}
\end{table}


\begin{figure*}[htp]
\includegraphics[width=1.0\textwidth]{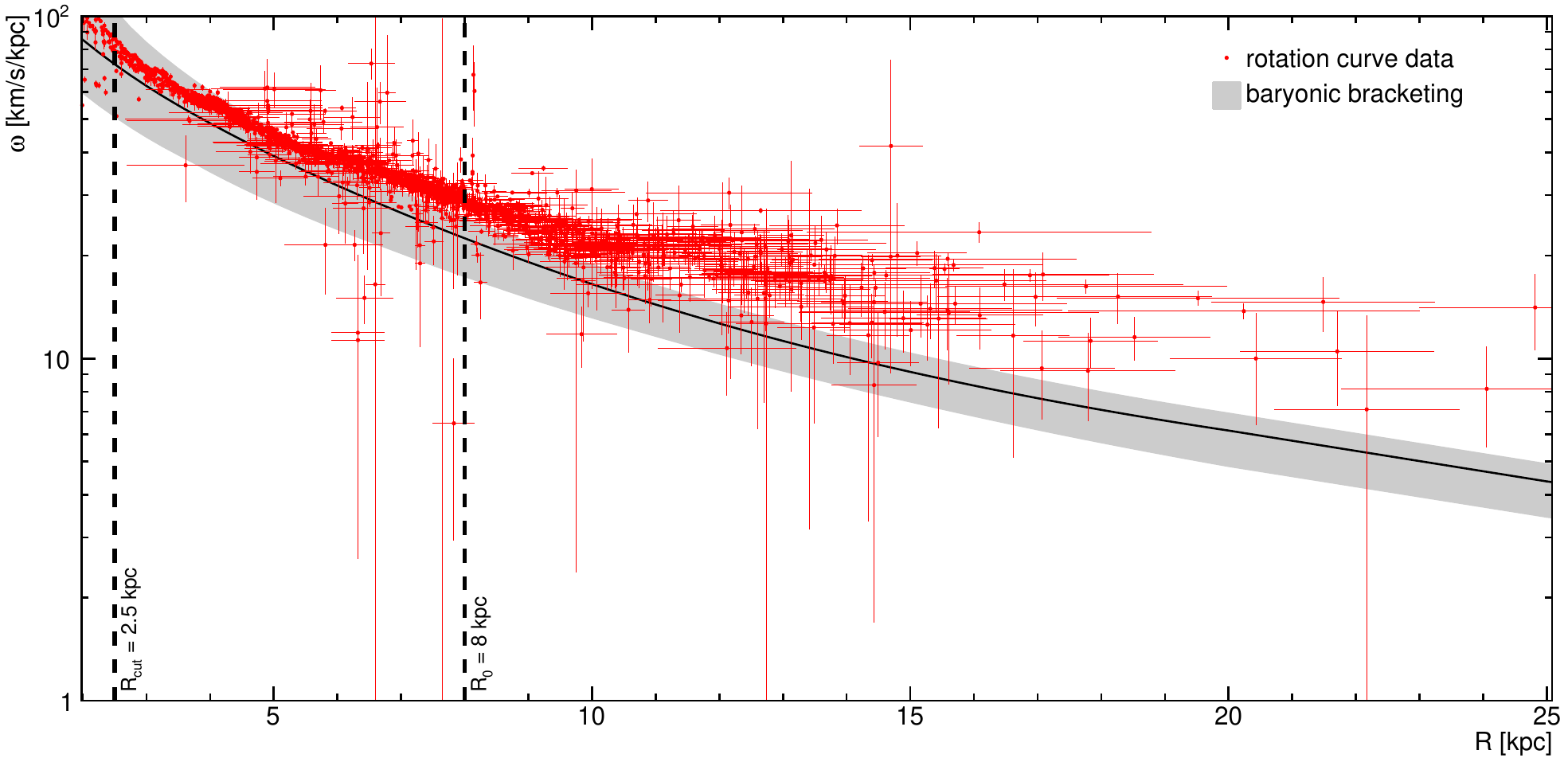}
\includegraphics[width=1.0\textwidth]{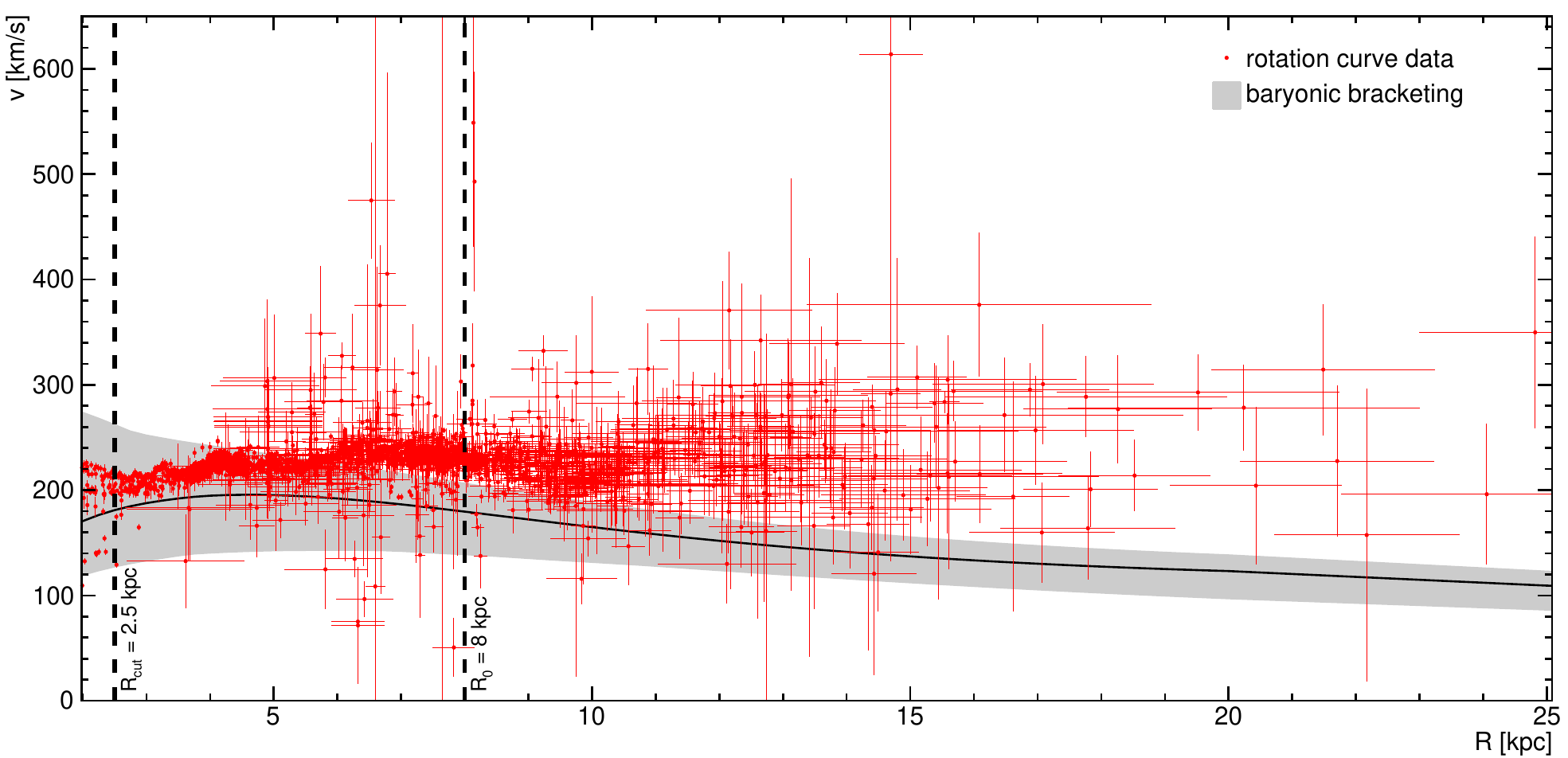}
\caption{The rotation curve of our Galaxy. The compilation of rotation curve measurements discussed in the text is shown by the red data points, whereas the bracketing of the 70 baryonic models implemented is spanned by the grey band. Both elements are plotted with $1\sigma$ uncertainties. The mean rotation curve predicted by a representative baryonic model (consisting of bulge 1 \cite{Stanek1997}, disc 5 \cite{Bovy:2013raa} and gas 1 \cite{Ferriere1998}, cf.~Tab.~\ref{tab:models}) is denoted by the black solid line. For convenience, the angular circular velocity is shown in the upper panel and the actual circular velocity in the lower panel. Here we take $R_0=8\,$kpc, $v_0=230\,$km/s and $\left(U,V,W\right)_{\odot}=(11.10,12.24,7.25)\,$km/s \cite{Schoenrich2010}.}
\label{fig:data}  
\end{figure*}

\begin{figure}[htp]
\centering
\includegraphics[width=0.49\textwidth]{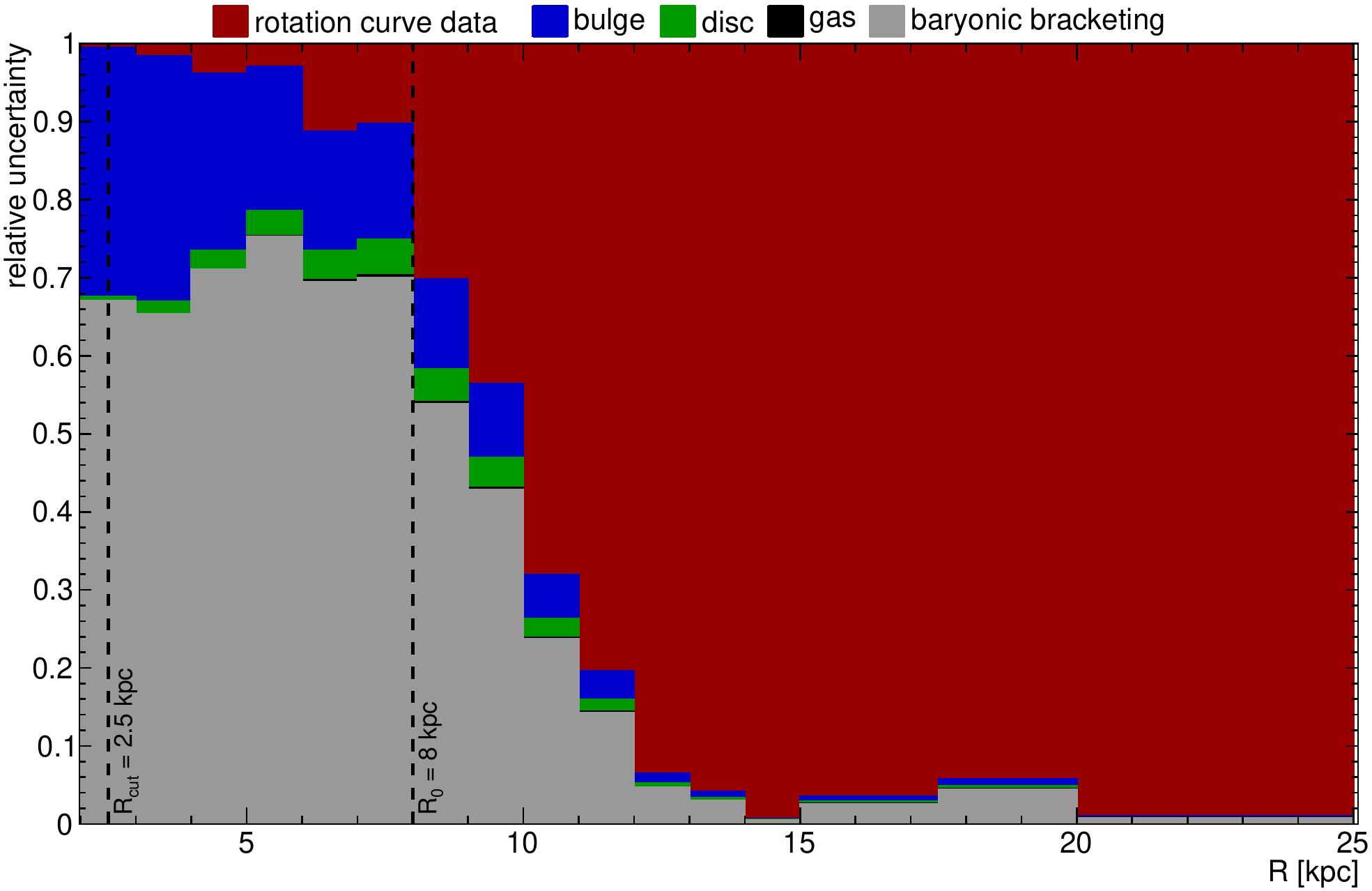}
\caption{Breakdown of the relative contribution to the overall uncertainty on the residual $\omega_c^2-\omega_\textrm{b}^2$ as a function of Galactocentric radius $R$. From top to bottom, the bars show the mean contribution due to rotation curve data, bulge, disc, gas and baryonic bracketing. The larger radial bins at $R\gtrsim 15\,$kpc are chosen due to the fewer $\omega_c$ measurements in that region. As in Fig.~\ref{fig:data}, we take $R_0=8\,$kpc, $v_0=230\,$km/s and $\left(U,V,W\right)_{\odot}=(11.10,12.24,7.25)\,$km/s \cite{Schoenrich2010}.}
\label{fig:errorbudget}  
\end{figure}

\par As for baryons, despite the improvements in the amount and quality of observations, their distribution in the Milky Way is still not known to high accuracy. In order to address the uncertainty on the density profile of stars and gas, we have carried out in Ref.~\cite{2015NatPh..11..245I} an extensive literature survey (cf.~Tab.~\ref{tab:models}) to bracket the allowed three-dimensional morphologies of each baryonic component, namely stellar bulge, stellar disc(s) and gas. Each component is now discussed in turn.
\par {\bf Stellar bulge.} The bulge dominates the inner $2-3\,$kpc of our Galaxy and presents a triaxial shape with a bar extending at positive Galactic longitudes. This general picture is consistently painted by different observations, but the morphological details are rather uncertain. We implement seven alternative data-based configurations for the bulge: exponential E2 \cite{Stanek1997}, gaussian G2 \cite{Stanek1997}, gaussian plus nucleus \cite{Zhao1996}, truncated power laws \cite{BissantzGerhard2002,Vanhollebeke2009}, power law plus long bar \cite{LopezCorredoira2007} and double ellipsoid \cite{Robin2012}. All models are normalised to the observed microlensing optical depth towards $(\ell,b)=(1.50^{\circ},-2.68^{\circ})$, $\langle \tau \rangle = 2.17^{+0.47}_{-0.38}\times 10^{-6}$ \cite{MACHO2005,Iocco2011}.
\par {\bf Stellar disc(s).} The structure of the Galactic disc has been thoroughly studied using different photometric surveys. The disc is usually modelled as a flattened component with a fastly decaying profile and is often subdivided into thin and thick populations. As with the bulge, there is no consensus over the global morphological details. We take therefore five alternative morphologies: pure thin plus thick discs \cite{HanGould2003,CalchiNovatiMancini2011}, thin plus thick discs with a stellar halo component \cite{deJong2010,Juric2008} and a single maximal disc \cite{Bovy:2013raa}. All models are normalised to the recent measurement of the local total stellar surface density, $\Sigma_{*}=38\pm 4\,\textrm{M}_\odot\textrm{/pc}^2$ \cite{Bovy:2013raa}.
\par {\bf Gas.} The gas component takes the form of molecular, atomic and ionised hydrogen (as well as a small fraction of heavier elements), and its distribution is fairly well-known. We model the gas in the inner $10\,$pc (effectively, a point-like distribution for our purposes) according to Ref.~\cite{Ferriere2012} and in the inner $2\,$kpc according to Ref.~\cite{Ferriere2007}. Above $2\,$kpc we implement two alternative morphologies based on Refs.~\cite{Ferriere1998,Moskalenko2002}. The uncertainties associated to each gas model stem mainly from the poorly constrained CO-to-H$_2$ conversion factor: $(2.5-10)\times 10^{19}\,\textrm{cm}^{-2}\textrm{(K km/s)}^{-1}$ for $R<2\,$kpc and $(0.5-3.0)\times 10^{20}\,\textrm{cm}^{-2}\textrm{(K km/s)}^{-1}$ for $R>2\,$kpc \cite{Ferriere1998,Ackermann2012}.
\par Tab.~\ref{tab:models} summarises all models specified above. For each bulge, disc and gas model, it is now possible to compute the corresponding contribution to the rotation curve. With seven bulges, five discs and two gas configurations, there are in total 70 different baryonic distributions, which we use to bracket the overall baryonic contribution $\omega_\textrm{b}^2 = \omega_\textrm{bulge}^2 + \omega_\textrm{disc}^2 + \omega_\textrm{gas}^2$.


\par In Fig.~\ref{fig:data} we show the rotation curve measurements and the expected baryonic contribution. The red data points show the compilation of kinematic tracers with 1$\sigma$ uncertainties. The grey band represents instead the envelope of the 70 baryonic models, including their corresponding 1$\sigma$ uncertainties, and the black solid line indicates the mean curve of a representative baryonic model defined by bulge 1, disc 5 and gas 1 (i.e.~E2 bulge \cite{Stanek1997}, single maximal disc \cite{Bovy:2013raa} and gas from Ref.~\cite{Ferriere1998}, cf.~Tab.~\ref{tab:models}). The upper and lower panels show, respectively, the angular circular velocity $\omega_c$, which will be used throughout our analysis, and the actual circular velocity $v_c$, which is the familiar quantity commonly displayed in rotation curve studies. Let us notice once again that the errors of $R$ and $v_c$ are strongly (positively) correlated, while those of $R$ and $\omega_c$ are uncorrelated. It is apparent from Fig.~\ref{fig:data} that baryons fall short of supporting the rotation curve in the Milky Way; a detailed analysis treating carefully statistical and systematic uncertainties shows the discrepancy is significant already inside the solar circle \cite{2015NatPh..11..245I}.

\par There are five sources of uncertainty that affect the comparison between $\omega_c$ and and $\omega_\textrm{b}$: the measurement of $\omega_c$ itself, the normalisations of bulge, disc and gas and the baryonic model bracketing. It is instructive to quantify the relative importance of each uncertainty on, for instance, the residual $\omega_c^2-\omega_\textrm{b}^2$. This is shown in Fig.~\ref{fig:errorbudget}. We find that for $R\lesssim R_0$ baryonic bracketing and, to a lesser extent, the bulge normalisation are the main sources of uncertainty, whereas for $R\gtrsim R_0$ the contribution of rotation curve measurements dominates over all other components. The uncertainties on disc and gas normalisations are small across the full radial range. In order to improve the dynamical constraints on the mass distribution in the Milky Way (including the ones shown later on in this work), it is therefore crucial to improve our knowledge of the spatial distribution of baryons at $R\lesssim R_0$ and to increase the precision of rotation curve data at $R\gtrsim R_0$. Fortunately, Gaia \cite{2012Ap&SS.341...31D}, APOGEE-2 (SDSS-IV) \cite{apogee2site}, WFIRST \cite{2015arXiv150303757S}, WEAVE \cite{weavesite} and 4MOST \cite{2012SPIE.8446E..0TD} should be able to help out on both fronts over the coming years.

\begin{figure*}[htp]
\centering
\includegraphics[width=0.49\textwidth]{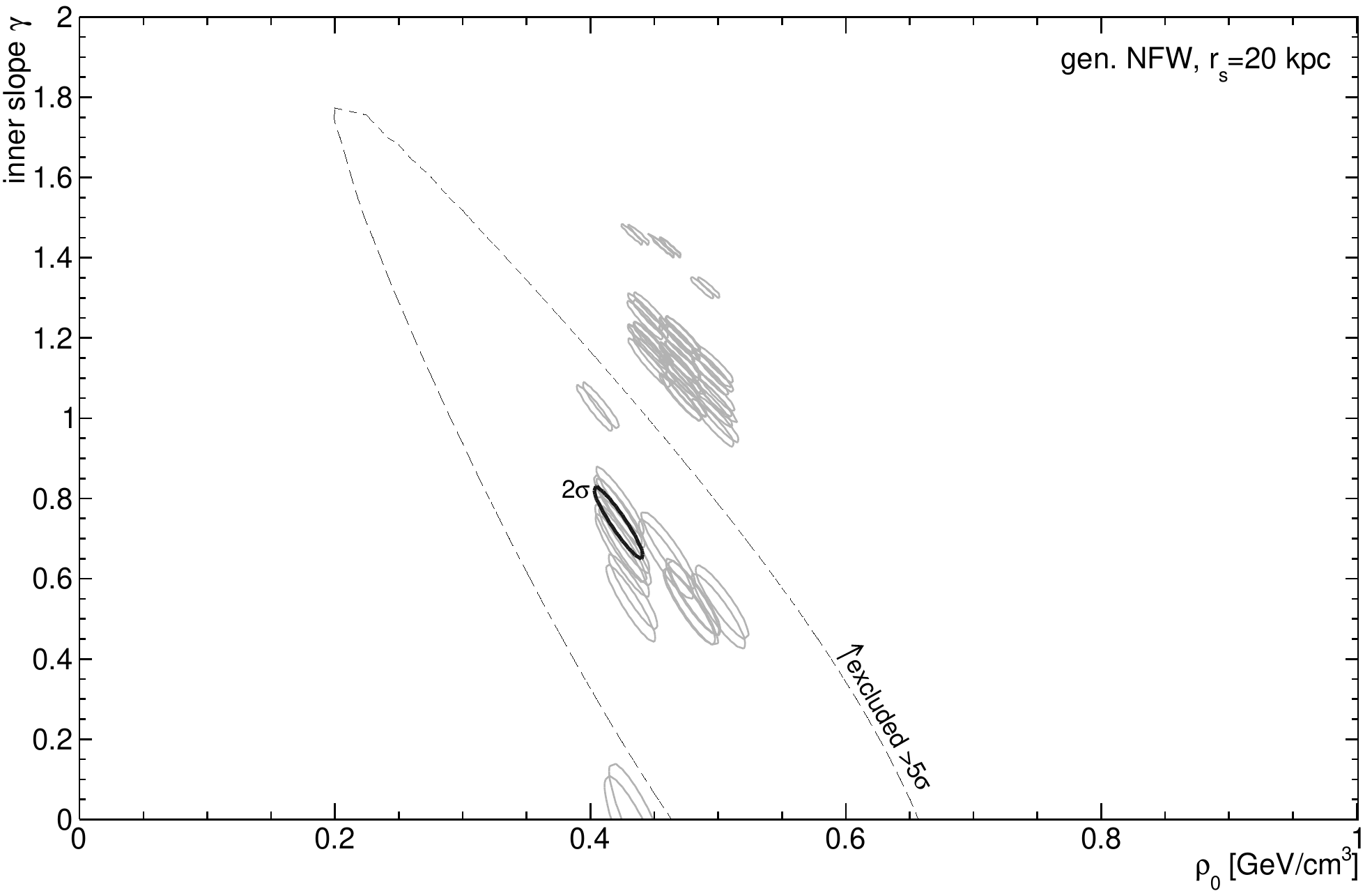}
\includegraphics[width=0.49\textwidth]{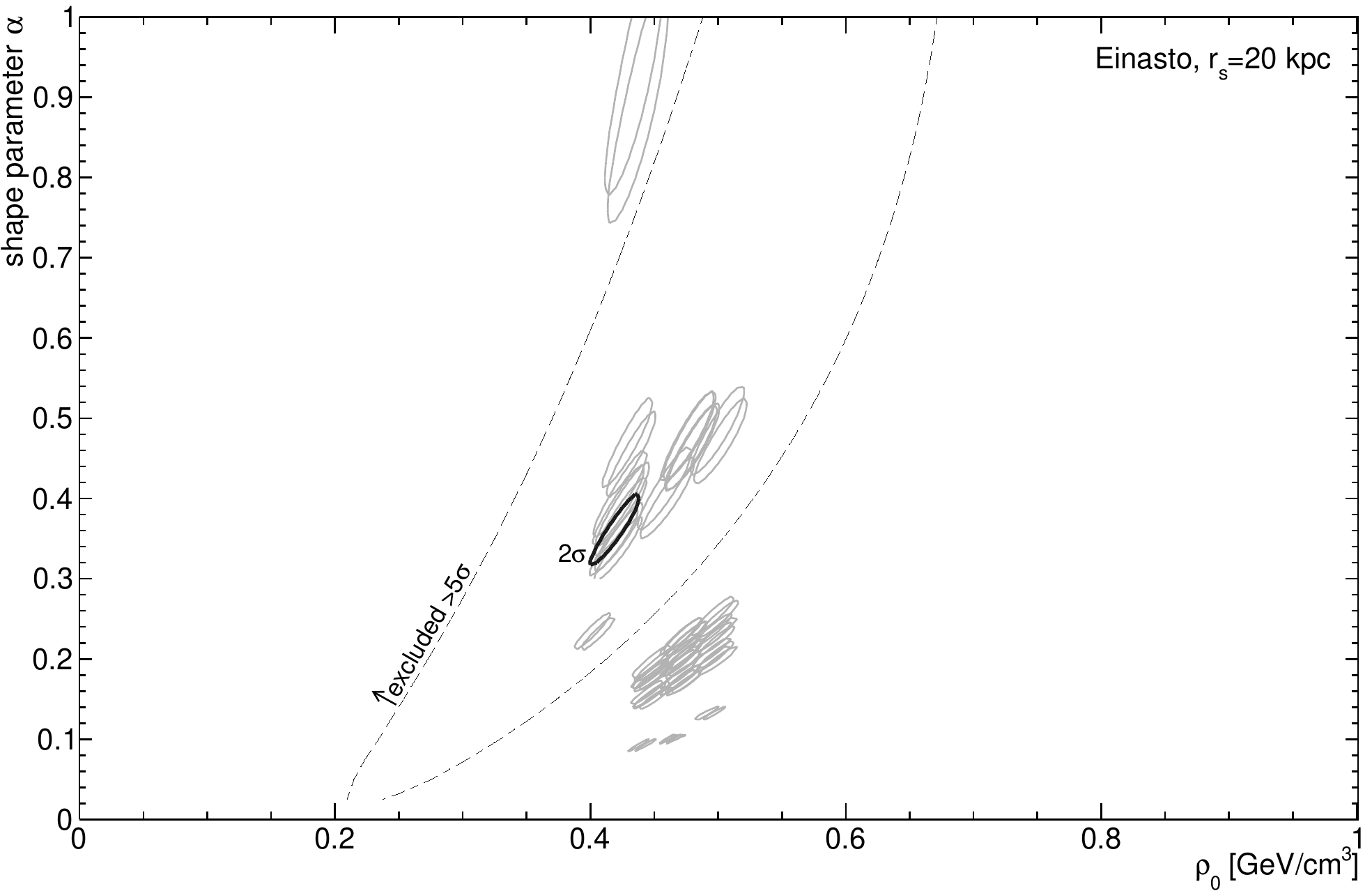}
\includegraphics[width=0.49\textwidth]{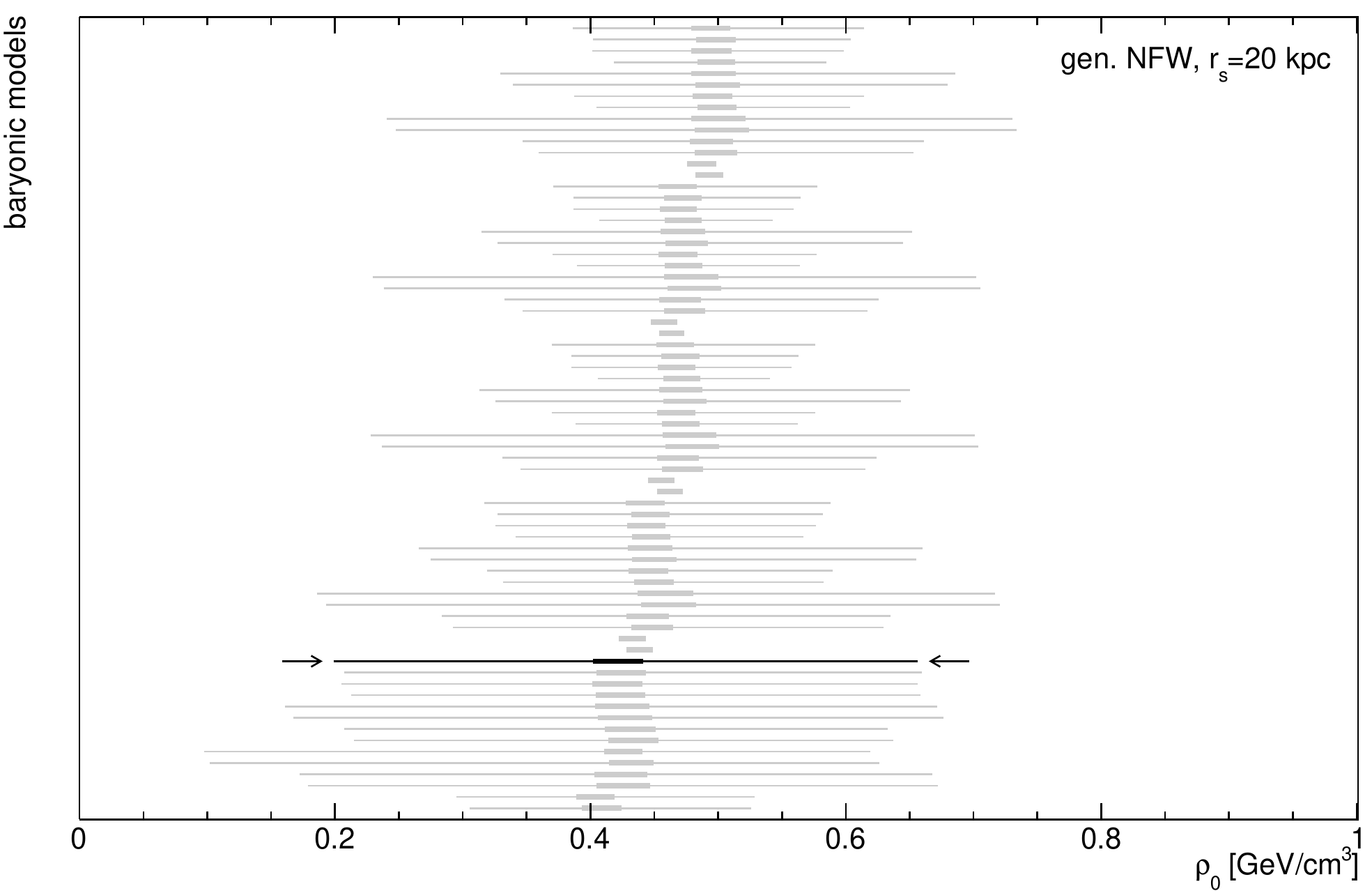}
\includegraphics[width=0.49\textwidth]{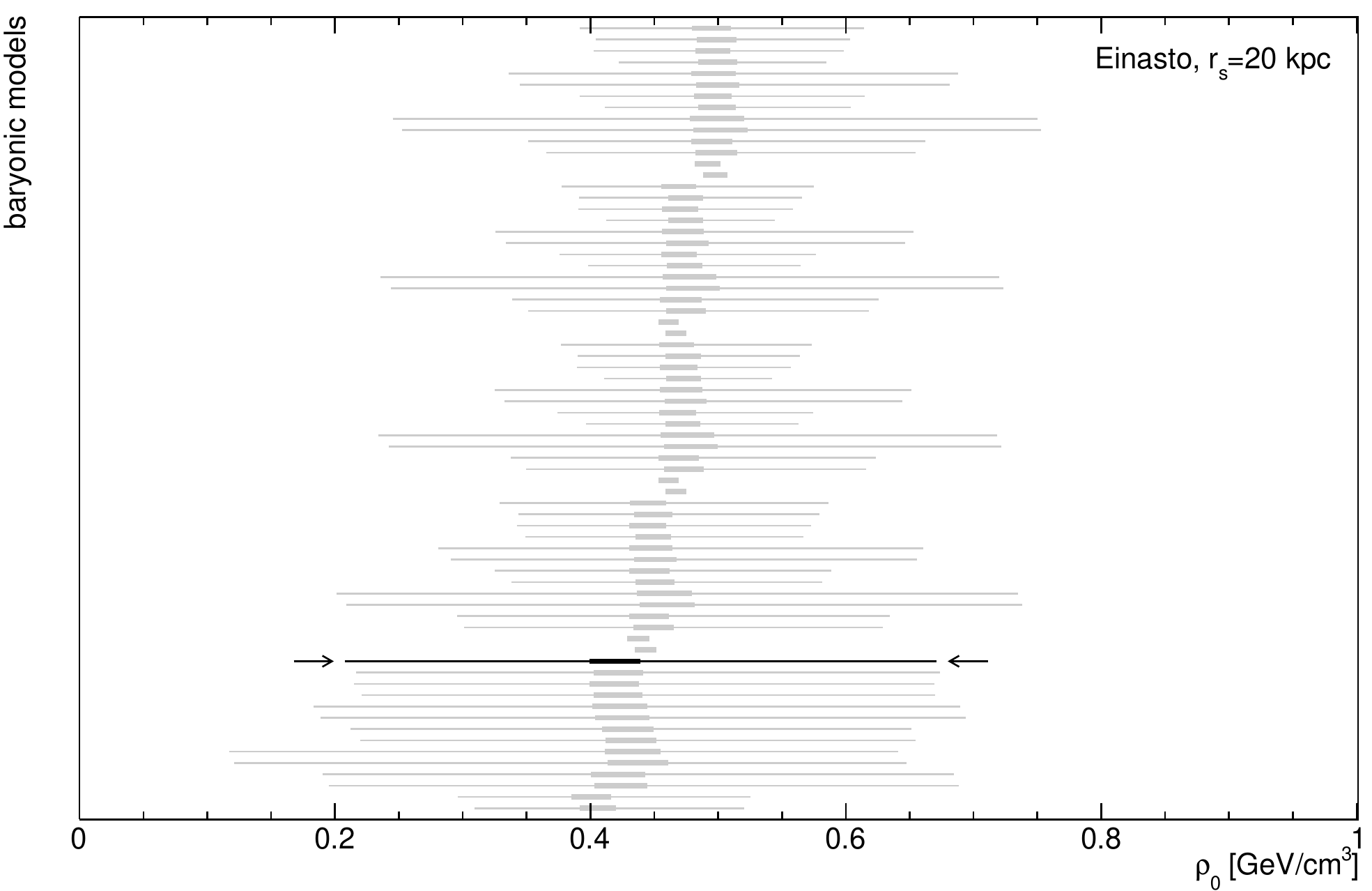}
\caption{Dark matter fit to the observed rotation curve of our Galaxy. The left (right) panels show the favoured regions of the dark matter parameter space for a generalised NFW (Einasto) profile with a fixed scale radius $r_s=20\,$kpc. The upper panels display with grey lines the $2\sigma$ confidence regions corresponding to each baryonic model. For the representative baryonic model \cite{Stanek1997,Bovy:2013raa,Ferriere1998}, besides the $2\sigma$ confidence region in thick black, we also plot the $5\sigma$ goodness-of-fit region in thin black dashed. The bottom panels show for each baryonic model the profiled local dark matter density range encompassed by the $2\sigma$ confidence regions (thick) and by the $5\sigma$ goodness-of-fit regions (thin). The baryonic models are ordered from top to bottom as in Tab.~\ref{tab:modelschi2}. The representative model is indicated by the arrows. This figure assumes $R_0=8\,$kpc, $v_0=230\,$km/s and $\left(U,V,W\right)_{\odot}=(11.10,12.24,7.25)\,$km/s \cite{Schoenrich2010}. For reference, $0.38\,$GeV/cm$^3 = 0.01\,$M$_\odot$/pc$^3$.}
\label{fig:DMfitting}  
\end{figure*}

\begin{figure}[htp]
\centering
\includegraphics[width=0.49\textwidth]{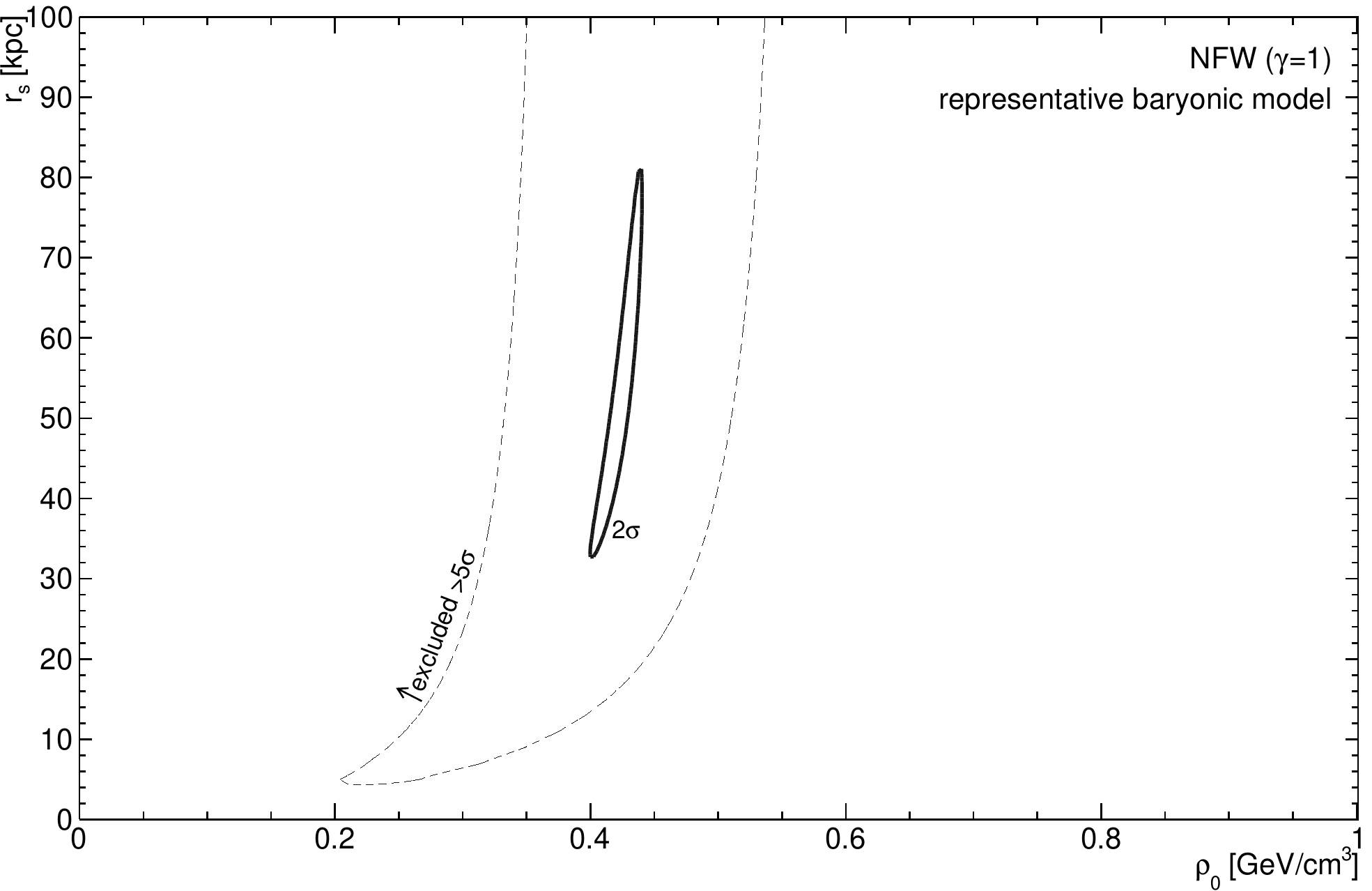}
\caption{The dependence of the dark matter fit on the scale radius. The plot shows the favoured regions of the dark matter parameter space for a pure NFW profile ($\gamma=1$) and the representative baryonic model \cite{Stanek1997,Bovy:2013raa,Ferriere1998}. The contours and line coding are the same as in Fig.~\ref{fig:DMfitting} (top left). We have assumed here $R_0=8\,$kpc, $v_0=230\,$km/s and $\left(U,V,W\right)_{\odot}=(11.10,12.24,7.25)\,$km/s \cite{Schoenrich2010}. For reference, $0.38\,$GeV/cm$^3 = 0.01\,$M$_\odot$/pc$^3$.}
\label{fig:DMfittingRs}  
\end{figure}

\begin{figure*}[htp]
\centering
\includegraphics[width=0.49\textwidth]{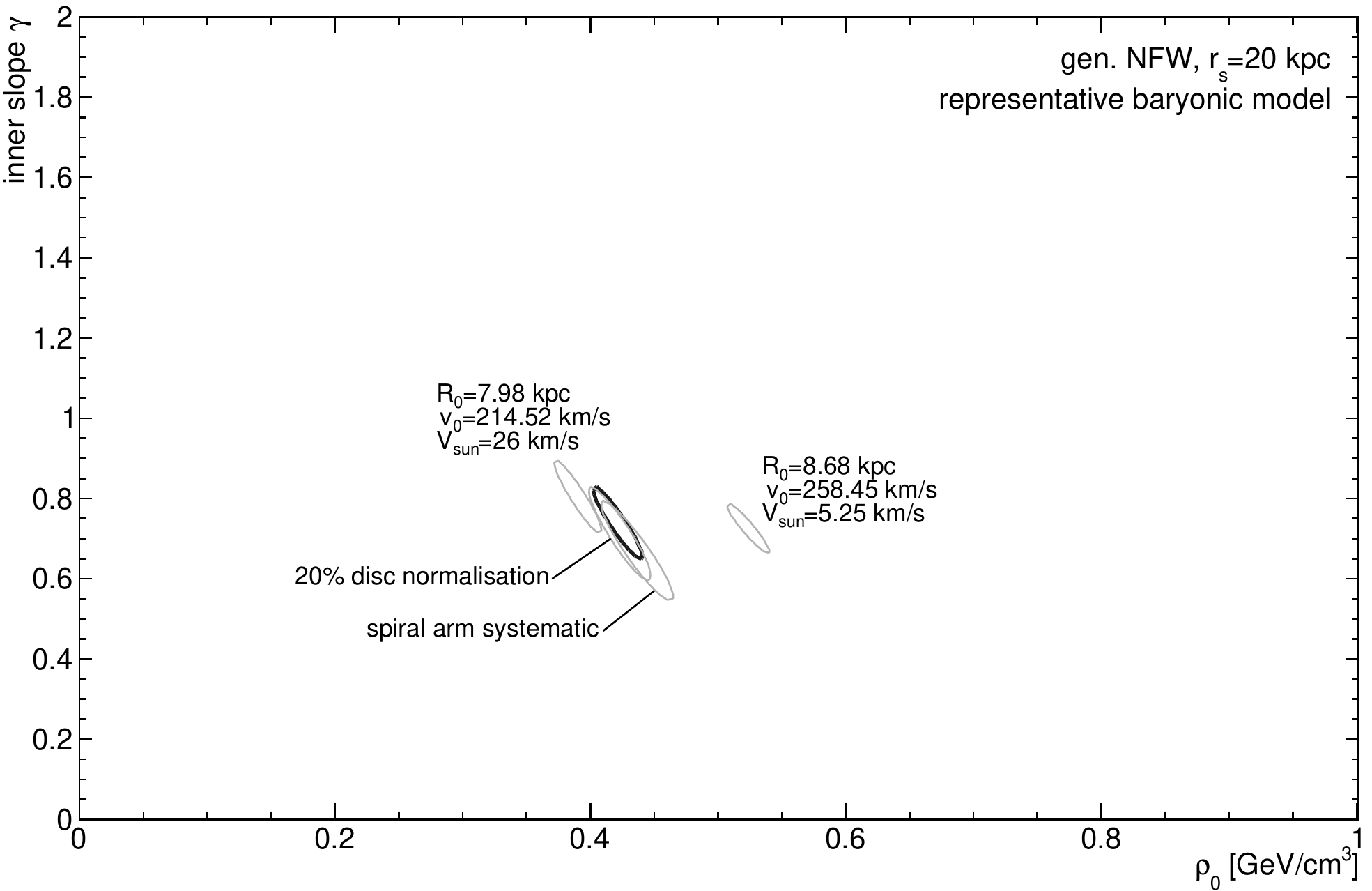}
\includegraphics[width=0.49\textwidth]{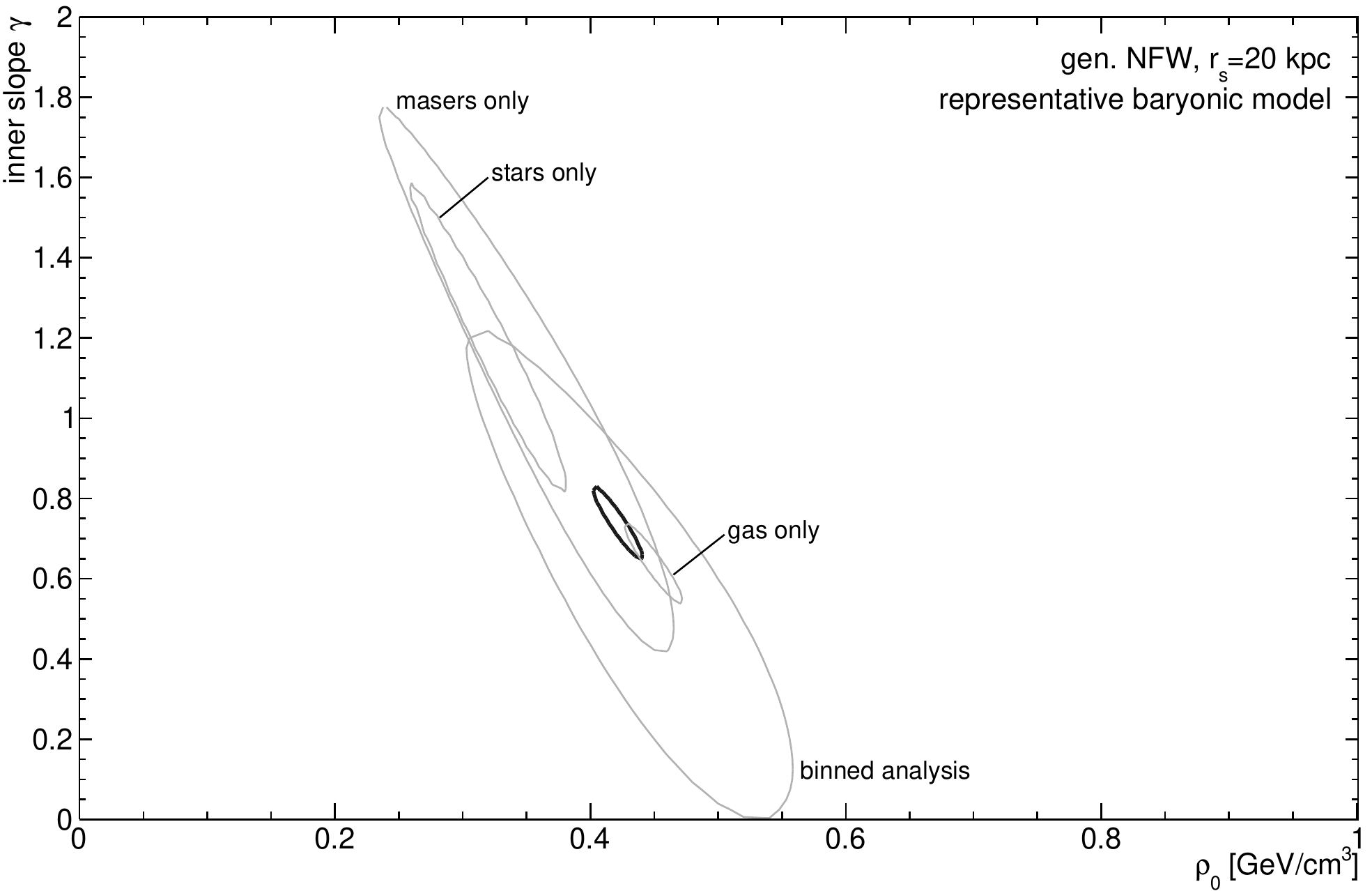}
\caption{The dependence of the dark matter fit on Galactic parameters, data selection and systematics. In the left panel we show the effect of different values of $R_0$, $v_0$ and $V_\odot$, of a systematic due to spiral arms and of an enhanced stellar disc normalisation uncertainty. The right panel illustrates the impact of a different data selection (gas, stars, masers) as well as the result of a standard binned analysis for comparison. In all cases the $2\sigma$ confidence region for the generalised NFW profile and the representative baryonic model \cite{Stanek1997,Bovy:2013raa,Ferriere1998} is plotted. The thick black contour is the baseline case presented in Fig.~\ref{fig:DMfitting} (top left). For reference, $0.38\,$GeV/cm$^3 = 0.01\,$M$_\odot$/pc$^3$.}\label{fig:DMfitting2}  
\end{figure*}

\section{Methodology}\label{secmethod}

\par Having introduced the observed rotation curve $\omega_c$ and the baryonic contribution $\omega_\textrm{b}$ in Sec.~\ref{secobs}, we now proceed to interpret their discrepancy in terms of dark matter. The goal here is to determine the best-fit parameters for a dark matter distribution by fitting the resulting total expected curve (baryons and dark matter) to the observational data. (For a different procedure to extract the dark matter profile directly from the data, see Ref.~\cite{2015ApJ...803L...3P}). We make the common assumption of a spherical dark matter distribution, for which the additional contribution to the rotation curve simply traces the total dark mass enclosed within a given radius $R$, namely
\begin{equation}\label{wdm2}
\omega_\textrm{dm}^2 = \frac{G M_{\textrm{dm}}(<\!\!R)}{R^3} = \frac{G}{R^3} \int_0^{R}{dr\, 4\pi r^2 \rho_\textrm{dm}(r)} \, .
\end{equation}
This then leads to a total angular velocity $\omega_t = \sqrt{w_\textrm{b}^2+\omega_\textrm{dm}^2}$, to be compared to the observed $\omega_c$. For the (spherical) dark matter density in Eq.~\eqref{wdm2}, we use either the generalised Navarro-Frenk-White (NFW) profile \cite{1990ApJ...356..359H,NFW1996,Navarro:1996gj,Merritt2006}, $\rho_\textrm{dm}\propto(r/r_s)^{-\gamma}(1+r/r_s)^{-3+\gamma}$, where $\gamma$ is the inner slope and $r_s$ is the scale radius, or the Einasto profile \cite{Navarro:2003ew,Merritt2006}, $\rho_\textrm{dm}\propto\exp(-2((r/r_s)^\alpha-1)/\alpha)$, where $\alpha$ is a shape parameter. In both cases, we parametrise the normalisation of the profiles in terms of the the local dark matter density $\rho_0 \equiv \rho_\textrm{dm}(R_0)$.

\par The remaining task is to fit $\omega_t$ to $\omega_c$ and derive constraints on the dark matter parameters $\rho_0$, $\gamma$ (or $\alpha$) and $r_s$. A subtlety here concerns the large errors on $R$ present in the compilation of kinematic data (cf.~Fig.~\ref{fig:data}), which preclude the use of a standard chi-square fitting. Instead of binning the data in $R$ (and thus necessarily losing information), we follow Ref.~\cite{Fich1989} to define a chi-square that accounts for uncertainties on both the vertical and horizontal axes. Defining the dimensionless variables $(x,y)=(R/R_0,\,\omega_c/\omega_0-1)$ with $\omega_0=v_0/R_0$, we have
\begin{equation}\label{chi2}
\chi^2 = \sum_{i=1}^{N} { d_i^2 } \equiv \sum_{i=1}^{N} { \left[ \frac{(y_i-y_{t,i})^2}{\sigma_{y,i}^2+\sigma_{\textrm{b},i}^2} + \frac{(x_i-x_{t,i})^2}{\sigma_{x,i}^2}  \right] } \, ,
\end{equation}
where $(x_i\pm\sigma_{x,i},y_i\pm\sigma_{y,i})$ are the rotation curve measurements discussed in Sec.~\ref{secobs}, $\sigma_{\textrm{b},i}$ is the uncertainty of the individual baryonic model evaluated at $x_i$  and $(x_{t,i},y_{t,i})$ are the points that minimise $d_i$ along the curve $y_t(x)=\omega_t(R\!=\!xR_0)/\omega_0-1$. The sum runs over all the $N$ objects in the compilation at $R>R_{\text{cut}}=2.5\,$kpc ($N=2687$ with 2081, 506 and 100 from gas, stars and masers, respectively) in order to exclude the innermost regions of the Galaxy where axisymmetry breaks down and some tracers may present non-circular orbits. We note in passing that Eq.~\eqref{chi2} can be applied since $x$ and $y$ (i.e.~$R$ and $\omega_c$) have uncorrelated errors. All results shown in the remainder of the work are based on the chi-square statistic defined above. We wish to point out that in Ref.~\cite{2015NatPh..11..245I} we have checked through Monte Carlo calculations that this statistic follows closely a chi-square distribution with the appropriate number of degrees of freedom for the case of our representative baryonic model and using the typical radial and velocity uncertainties found in our kinematic compilation.

\section{Results}\label{secDM}

\par We now turn to deriving constraints on the Galactic dark matter profile. For each of the 70 baryonic models described in Sec.~\ref{secobs}, a scan was performed over the dark matter profile parameter space and the $\chi^2$ computed as in Eq.~\eqref{chi2}. We first focus our attention on the parameter space $(\rho_0,\gamma)$ for NFW and $(\rho_0,\alpha)$ for Einasto ignoring the correlation between these parameters and other quantities such as $r_s$, $R_0$, $v_0$ or $V_\odot$, and then we explore quantitatively the effect of varying the scale radius, Galactic parameters, data selection and systematics.

\par Fig.~\ref{fig:DMfitting} presents the main results of our analysis for the generalised NFW (left) and Einasto (right) profiles, both with scale radius $r_s=20\,$kpc and for our baseline Galactic parameter configuration with $R_0=8\,$kpc, $v_0=230\,$km/s and $\left(U,V,W\right)_{\odot}=(11.10,12.24,7.25)\,$km/s \cite{Schoenrich2010}. We convey our results with two distinct regions in the parameter space: (i) the 95.45\% ($2\sigma$) confidence region encompassed by $\chi^2 \leq \chi^2_{\textrm{bf}}+\Delta\chi^2$, where $\chi^2_{\textrm{bf}}$ is the $\chi^2$ of the best fit configuration and $\Delta\chi^2=6.18$ (corresponding to two fitted parameters at $2\sigma$); and (ii) the $5\sigma$ goodness-of-fit region defined by $\chi^2\leq \chi^2_{5\sigma}$, where $\chi^2_{5\sigma}$ is the $\chi^2$ corresponding to an equivalent $5\sigma$ significance (i.e.~p-value $2.87\times10^{-7}$). The top panels display the $2\sigma$ confidence regions for all baryonic models and also the $5\sigma$ goodness-of-fit region for the representative baryonic model \cite{Stanek1997,Bovy:2013raa,Ferriere1998}. The bottom panels show instead the profiled range of local dark matter density encompassed by the $2\sigma$ confidence region and the $5\sigma$ goodness-of-fit region of all baryonic models. In Tab.~\ref{tab:modelschi2} we report the best fits for all baryonic models along with the profiled ranges of local dark matter density.

\par The constraints in Fig.~\ref{fig:DMfitting} are significantly more sensitive to the local dark matter density than to the slope of the profile. This simply reflects the impact of these two quantities on the dark matter content of the Galaxy and thus on its contribution to the rotation curve. Overall, the remarkably precise $2\sigma$ confidence regions arise from the choice of an unbinned analysis and the large number of kinematic tracers adopted. However, Fig.~\ref{fig:DMfitting} clearly shows how confidence regions corresponding to different baryonic models shift sizeably across the parameter space, a feature which reflects the importance of the baryonic contribution to the rotation curve of the inner Galaxy. In other words, the precision currently allowed by kinematic measurements is hindered by large systematics associated with the distribution of baryons. This highlights the reason for adopting a comprehensive collection of baryonic models in our analysis: whereas a single, fiducial model may well represent a useful benchmark, it cannot be used for accurate dynamical constraints unless one has an a priori expectation about the actual baryon morphology of the Galaxy. The spread of the dark matter constraints over the whole range of baryonic models is a measure of the systematic uncertainty due to baryonic modelling, which cannot be dealt with by averaging nor by marginalising over the entire range of allowed morphologies. We therefore present the constraints for all baryonic models instead of focussing on a single model with unknown systematics.

\par Another interesting message conveyed by Fig.~\ref{fig:DMfitting} is that dynamical constraints start to be able to resolve between different baryonic models. Let us focus on the left panels corresponding to the NFW profile (very similar considerations can be drawn for the Einasto profile). From the bottom left plot it is clear that there are five blocks of baryonic models clustered around slightly different local dark matter densities. These five blocks correspond precisely to the five discs implemented in our analysis (cf.~Tab.~\ref{tab:models}). The inferred $\rho_0$ ranges from around $0.50\,\textrm{GeV/cm}^3$ for models including disc 1 down to around $0.42\,\textrm{GeV/cm}^3$ for models with disc 5, and in between for the other discs 2-4. These figures are comparable to but somewhat higher than values typically found when applying global methods \cite{CatenaUllio2010,Weber:2009pt}. The inner slope is less well determined, but the confidence regions in the top left panel of Fig.~\ref{fig:DMfitting} are mainly centred around three values of the inner slope ($\gamma\sim 0.6, 1.2, 1.4$) with a few outliers. This configuration correlates with the bulge used in each baryonic model. When using any of the discs 1-4, the baryonic models with the bulges 1-4 or 6 point to $\gamma\sim1.2$, the ones with bulge 5 cluster around $\gamma\sim0.6$ and the ones with bulge 7 tend towards $\gamma\sim1.4$. The case of models with disc 5 is slightly different: the bulges 1-4 or 6 point now to $\gamma\sim0.6$, while bulge 5 prefers shallow profiles $\gamma\sim0$ (in this case cored profiles are possibly favoured over NFW or Einasto) and bulge 7 points to $\gamma\sim1$. From all this discussion, we learn that the local dark matter density is particularly sensitive to the morphology of the stellar disc, whereas the inner slope of the dark matter profile is mostly dependent on the bulge configuration. The gas plays a relatively minor role.

\par It is worth noticing that there are eight baryonic models for which the $5\sigma$ goodness-of-fit region is vanishing (cf.~bottom panels in Fig.~\ref{fig:DMfitting} and Tab.~\ref{tab:modelschi2}). This is valid for both the generalised NFW and the Einasto profiles. These eight baryonic models coupled to all the dark matter profile configurations tested fall short of supporting the observed rotation curve and are excluded at $5\sigma$ level or more. The common characteristic of the eight models is the use of bulge 7 coupled to any of the two gas configurations and any of the discs except disc 5. If one were to have independent evidence for the dark matter profiles tested in Fig.~\ref{fig:DMfitting}, then bulge 7 (i.e.~the double ellipsoid bulge \cite{Robin2012}) would be too light and disfavoured by current dynamical data unless it is associated with disc 5 (i.e.~the single maximal disc \cite{Bovy:2013raa}).

\par For the representative baryonic model \cite{Stanek1997,Bovy:2013raa,Ferriere1998} (highlighted in black in the panels of Fig.~\ref{fig:DMfitting}), the $2\sigma$ ranges for the local dark matter density read
\begin{eqnarray}
\textrm{NFW:} \, & 	  \rho_0 = 0.420^{+0.021}_{-0.018} \, (2\sigma) \pm 0.025 \,\textrm{GeV/cm}^3 \, ,\\
\textrm{Einasto:} \, & \rho_0 = 0.420^{+0.019}_{-0.021} \, (2\sigma) \pm 0.026 \,\textrm{GeV/cm}^3 \, ,
\end{eqnarray}
where the second error is the standard deviation of the best-fit values over all baryonic models, which we use here as a measure of the current systematic uncertainty due to baryonic modelling. The interested reader can find in Tab.~\ref{tab:modelschi2} the inferred ranges for all baryonic models. For the sake of completeness, let us also point out that all values reported assume a spherical dark matter profile; an oblate profile as suggested by numerous numerical simulations would lead to a higher local dark matter density by tens of percent \cite{2010PhRvD..82b3531P,2014JCAP...09..004B}.

\par We now comment on the impact of breaking the assumptions behind our main results shown in Fig.~\ref{fig:DMfitting}. The effect of letting free the scale radius is illustrated in Fig.~\ref{fig:DMfittingRs} for a pure NFW profile with inner slope $\gamma=1$ and the representative baryonic model \cite{Stanek1997,Bovy:2013raa,Ferriere1998}. For comparison, numerical simulations of Milky Way-like halos find scale radii in the range $r_s\simeq12-37\,$kpc \cite{2002ApJ...573..597K,2004PhRvL..93x1301P,2008MNRAS.391.1940M}. As the kinematic data used in the current paper are restricted to the inner Galaxy (namely, $R\lesssim20\,$kpc), the $5\sigma$ exclusion region is rather loose unless $r_s\lesssim 20\,$kpc. Better constraints on the scale radius require tracers in the outer Galaxy, as pursued e.g.~in Ref.~\cite{2010ApJ...720L.108G,Nesti:2013uwa}. Notwithstanding, Fig.~\ref{fig:DMfittingRs} clearly shows that the local dark matter density is well constrained with uncertainties comparable to the ones obtained in Fig.~\ref{fig:DMfitting} (left) where we fixed $r_s=20\,$kpc but varied $\gamma$.

\par Finally, Fig.~\ref{fig:DMfitting2} depicts the impact of Galactic parameters, data selection and systematics in our dynamical constraints. For this figure we only plot the $2\sigma$ confidence region for the generalised NFW profile and the representative baryonic model \cite{Stanek1997,Bovy:2013raa,Ferriere1998}. The left panel shows the effect of changing the Galactic fundamental parameters $R_0$, $v_0$ and $V_\odot$. The current ranges for these quantities read $R_0=8.0\pm0.5\,$kpc \cite{Gillessen2009,Ando2011,Malkin2012,Reid2014}, $v_0=230\pm20\,$km/s \cite{Reid2009,Bovy2009, McMillan2010, Bovy2012,Reid2014} and $V_{\odot}=5.25-26\,$km/s \cite{DehnenBinney1998,Schoenrich2010,Bovy2012,Reid2014}, but there are important correlations among the different values. Let us consider specifically the measurement $R_0=8.33\pm0.35\,$kpc \cite{Gillessen2009} from the monitoring of stellar orbits around the central supermassive black hole and $\Omega_{\odot}\equiv\frac{v_0+V_{\odot}}{R_0}=30.26\pm0.12\,\color{black}$km/s/kpc \cite{ReidBrunthaler2004} from the proper motion of Sgr A$^{\ast}$. Put together, the two measurements define four $1\sigma$ configurations depending on the adopted value of $V_\odot$:\\
(a) $R_0=7.98\,$kpc, $v_0=214.52\,\color{black}$km/s for $V_{\odot}=26\,$km/s;\\ 
(b) $R_0=7.98\,$kpc, $v_0=237.18\,\color{black}$km/s for $V_{\odot}=5.25\,$km/s;\\ 
(c) $R_0=8.68\,$kpc, $v_0=235.62\,\color{black}$km/s for $V_{\odot}=26\,$km/s; \\ 
(d) $R_0=8.68\,$kpc, $v_0=258.45\,\color{black}$km/s for $V_{\odot}=5.25\,$km/s.\\ 
In the left panel of Fig.~\ref{fig:DMfitting2}, we present the results for the two most extreme configurations, which turn out to be (a) and (d). Clearly, the current (correlated) uncertainties on $R_0$, $v_0$ and $V_\odot$ hinder the determination of the local dark matter density, while their effect on the determination of the inner slope is somewhat less important. Also shown in the left panel is the impact of the systematic motion due to spiral arms modelled according to Ref.~\cite{BrandBlitz1993} and of an enhanced 20\% uncertainty on the local total stellar surface density (see the supplementary information of Ref.~\cite{2015NatPh..11..245I} for further details). Our constraints are fairly robust in both cases. The right panel of Fig.~\ref{fig:DMfitting2} displays instead how our confidence regions shift selecting separately gas kinematics, star kinematics and masers. Although there is some dependence on data selection, we opt to show our main results including all data available in the compilation of kinematic measurements. For comparison, we also plot the results of a standard binned analysis applied to the full data set, which is less precise overall but more robust against data selection.

\section{Conclusion}\label{secconc}

\par In the cold dark matter paradigm, the innermost regions of the Milky Way are expected to harbour a significant amount of dark matter. Testing such expectation with the help of observations has historically been difficult due to the uncertainties on both the distribution of baryons and the rotation curve. We have tried here to overcome those difficulties by combining a comprehensive compilation of rotation curve measurements with state-of-the-art baryonic models. This allowed us to effectively subtract the contribution of baryons off the observed rotation curve, and to identify the favoured range of the parameters of different dark matter profiles. We believe the data and analysis techniques presented here will be useful for future dynamical studies extending beyond standard assumptions (e.g.~non-spherical dark matter profiles). At present, the precision allowed by kinematic measurements is entirely overshadowed by the current uncertainty on the fundamental Galactic parameters and on the morphology of baryons. In that respect, the forthcoming data from the Gaia mission and surveys such as APOGEE-2 (SDSS-IV), WFIRST, WEAVE and 4MOST will play a crucial role in shrinking these uncertainties and will hopefully open a precision era in the measurement of the dark matter distribution in the Galaxy.


\vspace{0.5cm}
{\it Acknowledgements.} We would like to thank Alessandro Cuoco, Miguel A.~S\'anchez-Conde, Thomas Schwetz and Gabrijela Zaharijas for useful discussions. M.~P.~acknowledges the support from Wenner-Gren Stiftelserna in Stockholm, F.~I.~from the Simons Foundation and FAPESP process 2014/11070-2, and G.~B.~from the European Research Council through the ERC Starting Grant {\it WIMPs Kairos}.


\bibliographystyle{apsrev}
\bibliography{dynconst}


\begin{table*}
\begin{center}
\fontsize{8}{8}\selectfont
\begin{tabular}{ |cc| ccc ccc c |ccc ccc c| }	
\cline{3-16}
 \multicolumn{2}{c|}{} & \multicolumn{7}{c|}{generalised NFW ($r_s=20\,\textrm{kpc}$)} & \multicolumn{7}{c|}{Einasto ($r_s=20\,\textrm{kpc}$)} \\ \hline
 \multicolumn{2}{|c|}{baryonic model} && $\chi^2_{\textrm{bf}}/N$ && \multicolumn{4}{c|}{$\rho_0$ [GeV/cm$^3$]} && $\chi^2_{\textrm{bf}}/N$ && \multicolumn{4}{c|}{$\rho_0$ [GeV/cm$^3$]}  \\
\hline
1  & [1-1-1] 			&& 1.016 && 0.479--0.509 && (0.386--0.614) &&& 1.012 && 0.479--0.510 && (0.392--0.614) &\\
2  & [1-1-2] 			&& 1.042 && 0.483--0.514 && (0.402--0.604) &&& 1.037 && 0.484--0.514 && (0.404--0.604) &\\
3  & [2-1-1] 			&& 1.044 && 0.479--0.511 && (0.401--0.599) &&& 1.040 && 0.482--0.510 && (0.402--0.599) &\\
4  & [2-1-2] 			&& 1.071 && 0.484--0.513 && (0.418--0.585) &&& 1.067 && 0.485--0.515 && (0.422--0.585) &\\
5  & [3-1-1] 			&& 0.899 && 0.479--0.514 && (0.329--0.686) &&& 0.893 && 0.479--0.514 && (0.336--0.688) &\\
6  & [3-1-2] 			&& 0.919 && 0.482--0.517 && (0.339--0.680) &&& 0.913 && 0.483--0.516 && (0.345--0.681) &\\
7  & [4-1-1] 			&& 1.020 && 0.480--0.511 && (0.387--0.614) &&& 1.015 && 0.481--0.511 && (0.391--0.615) &\\
8  & [4-1-2] 			&& 1.046 && 0.484--0.514 && (0.405--0.604) &&& 1.041 && 0.484--0.514 && (0.412--0.604) &\\
9  & [5-1-1] 			&& 0.744 && 0.479--0.522 && (0.240--0.730) &&& 0.737 && 0.478--0.521 && (0.245--0.750) &\\
10 & [5-1-2] 			&& 0.757 && 0.482--0.524 && (0.248--0.734) &&& 0.751 && 0.480--0.523 && (0.252--0.753) &\\
11 & [6-1-1] 			&& 0.940 && 0.478--0.512 && (0.347--0.661) &&& 0.934 && 0.479--0.511 && (0.351--0.662) &\\
12 & [6-1-2] 			&& 0.962 && 0.482--0.515 && (0.360--0.653) &&& 0.956 && 0.482--0.515 && (0.366--0.655) &\\
13 & [7-1-1] 			&& 1.359 && 0.476--0.499 && --		   &&& 1.361 && 0.482--0.502 && --	       &\\
14 & [7-1-2] 			&& 1.405 && 0.482--0.504 && --	           &&& 1.406 && 0.488--0.507 && --	       &\\

15 & [1-2-1] 			&& 1.031 && 0.453--0.483 && (0.371--0.578) &&& 1.028 && 0.456--0.483 && (0.378--0.575) &\\
16 & [1-2-2] 			&& 1.057 && 0.457--0.487 && (0.387--0.565) &&& 1.054 && 0.461--0.488 && (0.391--0.566) &\\
17 & [2-2-1] 			&& 1.061 && 0.454--0.483 && (0.387--0.559) &&& 1.059 && 0.456--0.484 && (0.390--0.558) &\\
18 & [2-2-2] 			&& 1.089 && 0.458--0.487 && (0.407--0.543) &&& 1.087 && 0.461--0.488 && (0.413--0.545) &\\
19 & [3-2-1] 			&& 0.914 && 0.455--0.490 && (0.315--0.652) &&& 0.910 && 0.456--0.489 && (0.326--0.653) &\\
20 & [3-2-2] 			&& 0.936 && 0.459--0.492 && (0.327--0.645) &&& 0.931 && 0.460--0.492 && (0.334--0.646) &\\
21 & [4-2-1] 			&& 1.034 && 0.453--0.484 && (0.370--0.577) &&& 1.031 && 0.456--0.484 && (0.376--0.576) &\\
22 & [4-2-2] 			&& 1.061 && 0.458--0.488 && (0.390--0.564) &&& 1.057 && 0.460--0.488 && (0.398--0.565) &\\
23 & [5-2-1] 			&& 0.762 && 0.458--0.500 && (0.229--0.702) &&& 0.756 && 0.457--0.499 && (0.235--0.720) &\\
24 & [5-2-2] 			&& 0.776 && 0.461--0.502 && (0.238--0.705) &&& 0.770 && 0.459--0.501 && (0.244--0.723) &\\
25 & [6-2-1] 			&& 0.956 && 0.454--0.486 && (0.332--0.626) &&& 0.952 && 0.455--0.487 && (0.339--0.625) &\\
26 & [6-2-2] 			&& 0.979 && 0.458--0.490 && (0.347--0.617) &&& 0.975 && 0.460--0.490 && (0.351--0.618) &\\
27 & [7-2-1] 			&& 1.382 && 0.447--0.468 && --	 	   &&& 1.386 && 0.453--0.469 && --	       &\\
28 & [7-2-2] 			&& 1.429 && 0.454--0.474 && --	 	   &&& 1.434 && 0.459--0.475 && --	       &\\

29 & [1-3-1] 			&& 1.031 && 0.452--0.481 && (0.370--0.576) &&& 1.029 && 0.454--0.481 && (0.377--0.573) &\\
30 & [1-3-2] 			&& 1.058 && 0.455--0.486 && (0.385--0.563) &&& 1.055 && 0.459--0.486 && (0.390--0.564) &\\
31 & [2-3-1] 			&& 1.062 && 0.453--0.482 && (0.385--0.557) &&& 1.060 && 0.454--0.484 && (0.389--0.557) &\\
32 & [2-3-2] 			&& 1.090 && 0.457--0.486 && (0.406--0.540) &&& 1.087 && 0.459--0.486 && (0.411--0.542) &\\
33 & [3-3-1] 			&& 0.915 && 0.454--0.488 && (0.313--0.650) &&& 0.910 && 0.455--0.488 && (0.325--0.651) &\\
34 & [3-3-2] 			&& 0.936 && 0.457--0.491 && (0.326--0.643) &&& 0.931 && 0.458--0.491 && (0.333--0.644) &\\
35 & [4-3-1] 			&& 1.035 && 0.452--0.482 && (0.370--0.576) &&& 1.031 && 0.454--0.483 && (0.374--0.574) &\\
36 & [4-3-2] 			&& 1.061 && 0.456--0.486 && (0.388--0.562) &&& 1.058 && 0.459--0.486 && (0.397--0.563) &\\
37 & [5-3-1] 			&& 0.762 && 0.456--0.498 && (0.228--0.701) &&& 0.756 && 0.455--0.497 && (0.234--0.719) &\\
38 & [5-3-2] 			&& 0.777 && 0.459--0.501 && (0.237--0.704) &&& 0.770 && 0.458--0.500 && (0.242--0.722) &\\
39 & [6-3-1] 			&& 0.956 && 0.452--0.485 && (0.331--0.624) &&& 0.952 && 0.453--0.485 && (0.338--0.623) &\\
40 & [6-3-2] 			&& 0.979 && 0.456--0.488 && (0.345--0.615) &&& 0.975 && 0.458--0.489 && (0.350--0.616) &\\
41 & [7-3-1] 			&& 1.383 && 0.445--0.466 && --		   &&& 1.387 && 0.453--0.469 && --	       &\\
42 & [7-3-2] 			&& 1.430 && 0.452--0.472 && -- 		   &&& 1.435 && 0.459--0.475 && --	       &\\

43 & [1-4-1] 			&& 0.960 && 0.428--0.458 && (0.317--0.588) &&& 0.959 && 0.431--0.459 && (0.329--0.586) &\\
44 & [1-4-2] 			&& 0.982 && 0.432--0.462 && (0.327--0.582) &&& 0.981 && 0.434--0.464 && (0.343--0.579) &\\
45 & [2-4-1] 			&& 0.985 && 0.429--0.459 && (0.326--0.576) &&& 0.984 && 0.431--0.459 && (0.342--0.573) &\\
46 & [2-4-2] 			&& 1.008 && 0.433--0.462 && (0.341--0.567) &&& 1.007 && 0.435--0.463 && (0.349--0.567) &\\
47 & [3-4-1] 			&& 0.844 && 0.429--0.464 && (0.266--0.660) &&& 0.842 && 0.430--0.464 && (0.281--0.661) &\\
48 & [3-4-2] 			&& 0.861 && 0.433--0.467 && (0.275--0.655) &&& 0.859 && 0.434--0.468 && (0.290--0.655) &\\
49 & [4-4-1] 			&& 0.966 && 0.430--0.461 && (0.319--0.590) &&& 0.964 && 0.431--0.462 && (0.325--0.589) &\\
50 & [4-4-2] 			&& 0.988 && 0.434--0.465 && (0.331--0.583) &&& 0.986 && 0.435--0.466 && (0.338--0.581) &\\
51 & [5-4-1] 			&& 0.689 && 0.437--0.481 && (0.186--0.717) &&& 0.684 && 0.436--0.480 && (0.201--0.735) &\\
52 & [5-4-2] 			&& 0.701 && 0.440--0.483 && (0.193--0.721) &&& 0.696 && 0.439--0.482 && (0.209--0.738) &\\
53 & [6-4-1] 			&& 0.884 && 0.428--0.462 && (0.284--0.635) &&& 0.882 && 0.431--0.462 && (0.295--0.634) &\\
54 & [6-4-2] 			&& 0.902 && 0.432--0.465 && (0.292--0.629) &&& 0.901 && 0.434--0.465 && (0.301--0.629) &\\
55 & [7-4-1] 			&& 1.289 && 0.422--0.444 && --       	   &&& 1.293 && 0.429--0.446 && --	       &\\
56 & [7-4-2] 			&& 1.327 && 0.428--0.449 && --      	   &&& 1.331 && 0.435--0.452 && --	       &\\

$\,\,\,$57$^\ast$ & [1-5-1] 	&& 0.773 && 0.402--0.441 && (0.199--0.656) &&& 0.772 && 0.399--0.439 && (0.208--0.671) &\\
58 & [1-5-2] 			&& 0.783 && 0.405--0.444 && (0.207--0.659) &&& 0.782 && 0.402--0.441 && (0.217--0.674) &\\
59 & [2-5-1] 			&& 0.785 && 0.402--0.441 && (0.205--0.656) &&& 0.784 && 0.399--0.438 && (0.215--0.669) &\\
60 & [2-5-2] 			&& 0.796 && 0.404--0.443 && (0.213--0.658) &&& 0.795 && 0.403--0.441 && (0.221--0.670) &\\
61 & [3-5-1] 			&& 0.690 && 0.403--0.446 && (0.161--0.672) &&& 0.689 && 0.401--0.445 && (0.183--0.689) &\\
62 & [3-5-2] 			&& 0.699 && 0.406--0.448 && (0.168--0.676) &&& 0.698 && 0.404--0.446 && (0.189--0.694) &\\
63 & [4-5-1] 			&& 0.784 && 0.411--0.451 && (0.207--0.633) &&& 0.782 && 0.409--0.449 && (0.212--0.651) &\\
64 & [4-5-2] 			&& 0.795 && 0.414--0.453 && (0.215--0.637) &&& 0.793 && 0.412--0.451 && (0.220--0.655) &\\
65 & [5-5-1] 			&& 0.576 && 0.411--0.441 && (0.098--0.619) &&& 0.576 && 0.411--0.455 && (0.117--0.641) &\\
66 & [5-5-2] 			&& 0.583 && 0.415--0.449 && (0.102--0.626) &&& 0.582 && 0.413--0.461 && (0.121--0.648) &\\
67 & [6-5-1] 			&& 0.714 && 0.403--0.445 && (0.173--0.668) &&& 0.713 && 0.401--0.443 && (0.191--0.685) &\\
68 & [6-5-2] 			&& 0.724 && 0.405--0.447 && (0.179--0.672) &&& 0.723 && 0.403--0.445 && (0.195--0.688) &\\
69 & [7-5-1] 			&& 1.007 && 0.389--0.419 && (0.295--0.529) &&& 1.008 && 0.385--0.416 && (0.296--0.525) &\\
70 & [7-5-2] 			&& 1.022 && 0.393--0.424 && (0.305--0.526) &&& 1.022 && 0.391--0.420 && (0.309--0.520) &\\
\hline
\end{tabular}
\caption{The best fit and local dark matter density inferred using generalised NFW and Einasto profiles with fixed scale radius $r_s=20\,\textrm{kpc}$ for all baryonic models. The baryonic models are specified by the configurations of bulge, disc and gas (cf.~Tab.~\ref{tab:models}) in the form [b-d-g], where b,d,g stand for bulge, disc and gas, respectively. Besides the best fit $\chi^2$ for each model, we also report the profiled ranges of local dark matter density covered by the $2\sigma$ confidence region (no parentheses) and by the $5\sigma$ goodness-of-fit region (between parentheses) shown in the bottom panels of Fig.~\ref{fig:DMfitting}. The representative baryonic model is indicated with an asterisk. Here we take $R_0=8\,$kpc, $v_0=230\,$km/s and $\left(U,V,W\right)_{\odot}=(11.10,12.24,7.25)\,$km/s \cite{Schoenrich2010}. For reference, $0.38\,$GeV/cm$^3 = 0.01\,$M$_\odot$/pc$^3$.}\label{tab:modelschi2}
\end{center}
\end{table*}

\end{document}